\documentclass[11pt]{article}
\usepackage[dvips]{epsfig}
\newtheorem{definition_int}{Definition}
{\end{rm}\end{definition_int}}

\begin{document}
\huge
\noindent
\centerline{\bf Algorithms Visualization Tool for Students and }\\
\centerline{\bf Lectures in Computer Science}
\normalsize
\\
\\
\begin{center}
\noindent {\bf  Kaninda Musumbu} \\
Labri ~ UMR 5800 du CNRS\\
 Universit\'e Bordeaux 1, 
\\ 351, cours de la Lib\'eration, F-33.405 TALENCE (France) \\
Email: musumbu@labri.fr 
\end{center}

\paragraph{Abstract:}
The best way to understand complex data structures or algorithm  is to see 
them in action.  The present work presents a new tool, especially 
useful for students and lecturers in computer science. 
It is written in Java and developed at Bordeaux University of Sciences and 
Technology. Its purposes is to  help students in understanding classical 
algorithms by illustrating them in different ways: graphical (animated), 
formal, and descriptive. We think that it  can be useful to everyone interested 
in algorithms, in particular to students in computer science 
that want to beef up their readings and university lecturers in their
major effort to enhance the data structures and algorithms course.
 The main new thing of this tool is the fact of making it possible to the user 
to animate their own algorithms.


\section{Introduction}
The algorithmic  is a fundamental field in data processing, and very studied in the various university formations
. However, certain concepts are not always obvious to seize, and the teachers do not have tools to allow them to illustrate their remarks.
Though Al Khwarizmi's algorism referred only to the rules of performing 
arithmetic using Hindu-Arabic numerals, a partial formalization of what would 
become the modern algorithm began with attempts to solve the 
"decision problem" posed by David Hilbert in 1928.
 Subsequent formalizations were framed as attempts to define "effective 
calculability" or "effective method" those formalizations included the 
G\"odel$-$Herbrand$-$Kleene recursive functions of 1930, 1934 and 1935, 
Alonzo Church's lambda calculus of 1936, Emil Post's "Formulation 1" of 1936, 
and Alan Turing's Turing machines of 1936–7 and 1939. Giving a formal 
definition of algorithms, corresponding to the intuitive notion, remains a 
challenging problem. The analysis and study of algorithms is a discipline of
 computer science, and is often practiced abstractly without the use of a 
specific programming language or implementation. In this sense, algorithm
 analysis resembles other mathematical disciplines in that it focuses on the
 underlying properties of the algorithm and not on the specifics of any 
particular implementation. Usually {\it  pseudocode} is used for analysis as
 it is the simplest and most general representation. 
Knuth advises the reader that "the best way to learn an algorithm is to try it
 $\ldots$ immediately take pen and paper and work through an example". 
And besides a lot of people, instinctively, to understand an algorithm (to 
understand how it works) impossible by  a simple blow of eye.
However, this is not because many people who try to create such applications
 that we are found our happiness. 
A good number of them are really very interesting
 and excellent from a pedagogical point of view, but none is  sufficiently complete to can served as teaching tools.
And yet such a program would be very useful for the students having difficulties  to assimilate precise algorithms, but also for the professors who could use 
it and improve their courses while making use of it in real time like support.

 Our program is thus part of this lineage, and allows you to assist the understanding of algorithms. The user would have two information to enter: the data structure to view, and the algorithm to be used.
The objective of our paper is to discuss a number of important issues for
the design of Vizualisation tool  and
to contribute to improve the training of the algorithmic filed.

\section{ Motivation} 
The main objective is the construction of educational softwares adapted to 
learning.
 First and foremost, it is essential to have in mind that our 
application is designed for two very specific public: on the one side the 
lectures, who will use it as a teaching tool, and another side the students, 
for learning purposes. Perhaps, animating algorithms , on a screen, is very useful to check our self that what we want is indeed what we wrote! Quite often, by doing this, we discover that our algorithm is not accurate enough.

Animation is a complement of analysis, it is very interesting for students in 
"auto-formation" and in its understanding of the problem and  the development of their algorithmic solution, on the one hand and on the other hand, management of synchronization and visualization of the algorithms,

\section{Technical challenge}
The definition of a high level, simple programming language which would make
 it possible to the users to write their own algorithms and to test the 
classical ones. 
  Once an algorithm is ready, it can be run in many modes like step-by-step
 or breakpoint mode, so that the user has control over what he has to see, 
so that he understands it better. 

\subsection{Data structures managements}
The animation of programs through the use of real-time graphic displays
 will depend on the model's execution of the algorithm which will be chosen 
by the user.
Currently, our software treats all kind of binary trees. In the long term 
we hope to treat the graphs and finite-state machines.

\begin{figure}[htbp]
\centering{\includegraphics[width=5cm]{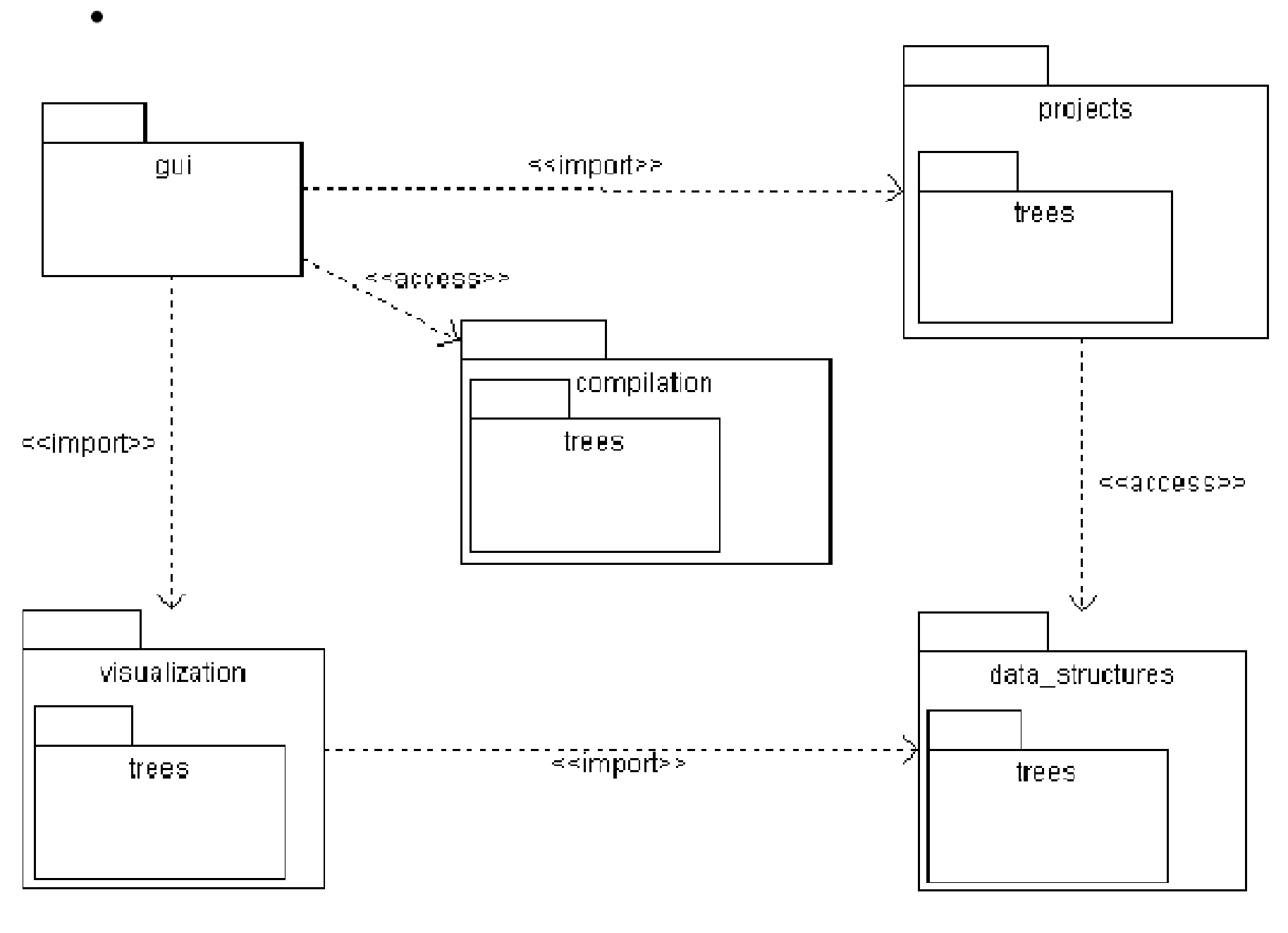 }}
 \caption{Global view of the architecture approach}
\label{fig:archi}
\end{figure}

\subsection{Compilation}
A compiler is a computer program that transforms source code written in (the source language) into another computer language (the target).
The preset algorithms as those of the user must be written in an easily comprehensible language as well by a student as by a professor.
A source text is known as syntactically correct if there exists a succession of
 derivations making it possible to build it starting from the axiom of a formal 
grammar. This sequence of derivations is a description of the text,
 called: {\it syntactic tree}. To avoid expense, we use a synthetic representation (of this syntactic tree) which contens only useful informations. 
This representation is an abstraction of the syntatic one: it is called
{\it  abstract tree}. This tree is used as support with the semantics analyzer, which visits the part {\it Instructions} (rules of derivations) to check of it semantics and also to generate the code targets corresponding.   
 The Declarations part is useful to produce the mechanism of identification. In a compiler, it is a table of the symbols which must be built in complement of the abstracted tree, to memorize the information attached to the declarations of a program. 
 Also, a new high level algorithmic language will have to be developed. Thus, during the execution of our various algorithms, our compiler must be able
 to detect any lexical, syntactic or semantic error and to announce it to the user. A successful compilation means that the algorithm was indeed translated into a programming language 

\subsection{Design of Interface : multiview editing }
The principal window must give us a sight (called “the scene”) on the chart 
of the data structure and another sight (zone of text) on the algorithm to be 
carried out. This simultaneous posting will make it possible to follow the 
graphic evolution of the algorithm carried out, which is essential on pedagogical point of view. and is more  efficient that the classical approach to debugging, (which consistes  to go through a program line by line and trace it, by 
watching every line, and the changes it makes in a window containing information about the contents of each variable or data-\ structure).   
This graphic interface would be complete with a menu for a better navigation in the various features of our application, and the tools facilitating the drafting and the exécution of algorithms. 

\section{Visualisation and  synchronisation}
For visualization we chose a management by frame. A frame is an object 
containing: a number of steps, the line to be highlighted, the list of the
 nodes roots, and the node currently selected. 
The recovery of the data we chose to use  a pattern: 
{\it  Observer/Observable.} 
We worked over again the compiler so that this one not provides us only an
 evaluation function  for the whole program but an evaluation function related 
to each instruction. We thus obtained an evaluation function  of the condition
{\it if..then...else}, the evaluation of an assignment, etc 
Thanks to this process, let us succeeded in recovering the data at the 
convenient period, inserting them in the frames and terminate the visualization
 and synchronization. 
The interface is a very important element, especially, as in our case, for a 
software which people will choose to use (it is not imposed), and which must 
thus allure its users. We especially wanted to give it an appearance closer to
 that of the applications of the operating system, and adapted according to the
 platform: Linux (Gnome), Windows… That very allows the user to
quickly find his brands. 
An important part of work on accessibility is the translation of the 
application in several languages. It is indeed an improvement which will 
touch a large audience. If the application is not represented in the language
 of the user,  
then it will not be used in most case. 
Documentation must be clear, simply, accessible, fast and to contain examples 
and assistances for the initial users, we have thus to create a documentation 
 online. 

\section{Conclusion and Perspectives }
Software visualization is  animated 2-D or 3-D visual representation of 
information about software systems based on their structure or behavior.
But it is not because many people test themselves has to create such applications that one finds our happiness. A good amount of between them are really very interesting and excellent from a teaching point of view, but none is sufficiently complete at the pedagagical point of view.
The creation of the language {\em jAlgo} was a scientific experiment and technological enriching. 
We thus propose a language {\it jAlgo} able to manage the recursivity, the algorithmic loops, conditions and other structures all via the call of functions.
 we did not implement the whole of the options and taken into account all the extensions of the software Nevertheless, the primary education and paramount objectives were reached we present the limits which we find with the product and the possible extensions of this last. 
We will like to be able to implement an evolutionary interface which would allow the user to visualize for each node, the  affected values.
With our taste the instructions suggested were not representative of all the situations with the which future users will have to cope. We would to have 
additional tools to traverse the binary trees with the algorithms assistance. 

\small
\bibliographystyle{plain}
\bibliography{Biblio}

\end{document}